\documentclass[twocolumn]{aastex62}

\graphicspath{{./}{figures/}}

\shorttitle{In-Situ Star Formation in the Outskirts of the LMC}
\shortauthors{Casetti-Dinescu et al.}

\begin{document}

\title{In-Situ Star Formation in the Outskirts of the Large Magellanic Cloud: Gaia DR2 Confirmation}

\correspondingauthor{Dana I. Casetti-Dinescu}
\email{dana.casetti@gmail.com} 


\author{Dana I. Casetti-Dinescu}
\affiliation{Physics Department, Southern Connecticut State University,
501 Crescent Street, New Haven CT06515, USA}
\affiliation{Astronomical Institute of the Romanian Academy, str. Cutitul de Argint 5, Bucharest, Romania}
\affiliation{Radiology and Biomedical Imaging, Yale School of Medicine, 300 Cedar Street, New Haven 06519, USA}
\author{Terrence M. Girard}
\affiliation{14 Dunn Road, Hamden, CT 06518, USA}
\author{Christian Moni Bidin}
\affiliation{Instituto de Astronom\'{i}a, Universidad Cat\'{o}lica del Norte, Av. Angomos 0610, Antofagasta, Chile}
\author{Lan Zhang}
\affiliation{Key Lab of Optical Astronomy, National Astronomical Observatories, CAS, 20A Datun Road, Chaoyang District, 100012 Beijing, China}
\author{Rene A. Mendez}
\affiliation{Departamento de Astronomia Universidad de Chile, Camino El observatorio \#1515, Las Condes, Santiago, Chile}
\author{Katherine Vieira}
\affiliation{Centro de Investigaciones de Astronomi\'{a}, Apartado Postal 264, M\'{e}rida 5101-A, Venezuela}
\author{Vladimir I. Korchagin}
\affiliation{Institute of Physics, Southern Federal University, Stachki st. 194, 344090, Rostov-on-Don, Russia}
\author{William F. van Altena}
\affiliation{Astronomy Department, Yale University, 46 Hillhouse Ave., New Haven, CT 06511, USA}

\begin{abstract}

We explore the Gaia DR2 proper motions of six young, main-sequence stars, members of the Large Magellanic Cloud (LMC) reported by \citet{mb17}. These stars are located in the outskirts of the disk, between $7\arcdeg$ and $13\arcdeg$ from the LMC's center where there is very low H~I content. Gaia DR2 proper motions confirm that four stars formed locally, in situ, while two are consistent with being expelled via dynamical interactions from inner, more gas-rich regions of the LMC. This finding establishes that recent star formation occurred in the periphery of the LMC, where thus far only old populations were known. 
\end{abstract}


\defcitealias{mb17}{MB17}

\section{Introduction} \label{sec:intro}
In a recent contribution, \citet[][hereafter MB17]{mb17} presented a spectroscopic analysis of a set of candidate young, B-type, main-sequence stars in the outskirts of the LMC. The candidates were selected via a large-area study that combined UV, optical and IR photometry to specifically look for young stars far from known regions of star formation \citep{cas12}. \citetalias{mb17} found six stars with distances and radial velocities consistent with LMC membership. They argued for in-situ star formation based on small line-of-sight velocity residuals from a disk model of the LMC. Lacking proper-motion measures and the remaining two velocity components, this result could not be conclusive. Indeed, \citet{boub17} proposed that these stars are runaways from the inner regions of the LMC. We revisit this issue using Gaia DR2 proper motions \citep{2016A&A...595A..1G, 2018A&A...}. Besides proper motions for our target stars, Gaia DR2 provides
proper motions across the full field of the LMC to an unprecedented combination of precision
and density. This allows us to work differentially, i.e., by obtaining proper-motion differences with respect to the local LMC motion within each subfield, without the need for a disk model. Moreover, we have a test case to interpret the kinematics, since one of our six stars is a known member of a young stellar association.
\section{Analysis} \label{sec:ana}
We extract the proper motions for our six young stars from the Gaia DR2 catalog \citep{2018A&A...}. Our stars have $G$ magnitudes between $\sim 15.1$ and 16.4, and thus are well measured by Gaia. The stars' IDs (Gaia DR2 and \citetalias{mb17}), 
proper motions and uncertainties are listed in Table \ref{tab1}, columns 1, 2, 7 and 8\footnote{Throughout the paper, proper-motion units are mas~yr$^{-1}$, and $\mu_{\alpha}$ is actually $\mu_{\alpha}$cos$\delta$.} respectively.
In addition, Gaia DR2 parallaxes confirm that these are distant stars, in agreement with the spectroscopic distances determined by \citetalias{mb17}. Specifically, all six stars have parallaxes compatible with zero at the $\pm 1\sigma_{\pi}$ level.

The bulk motion of the LMC, as well as its rotation --- nicely evidenced by \citet{helmi18} --- must be taken into account when analyzing the proper motions of our target stars. 
We first determine the mean motion of LMC stars in the area around each of our target stars. This local field motion is then subtracted from the proper motion of the target star to obtain the star's motion with respect to its LMC neighbors.

\begin{figure}
\includegraphics[width=0.3\textwidth,angle=-90]{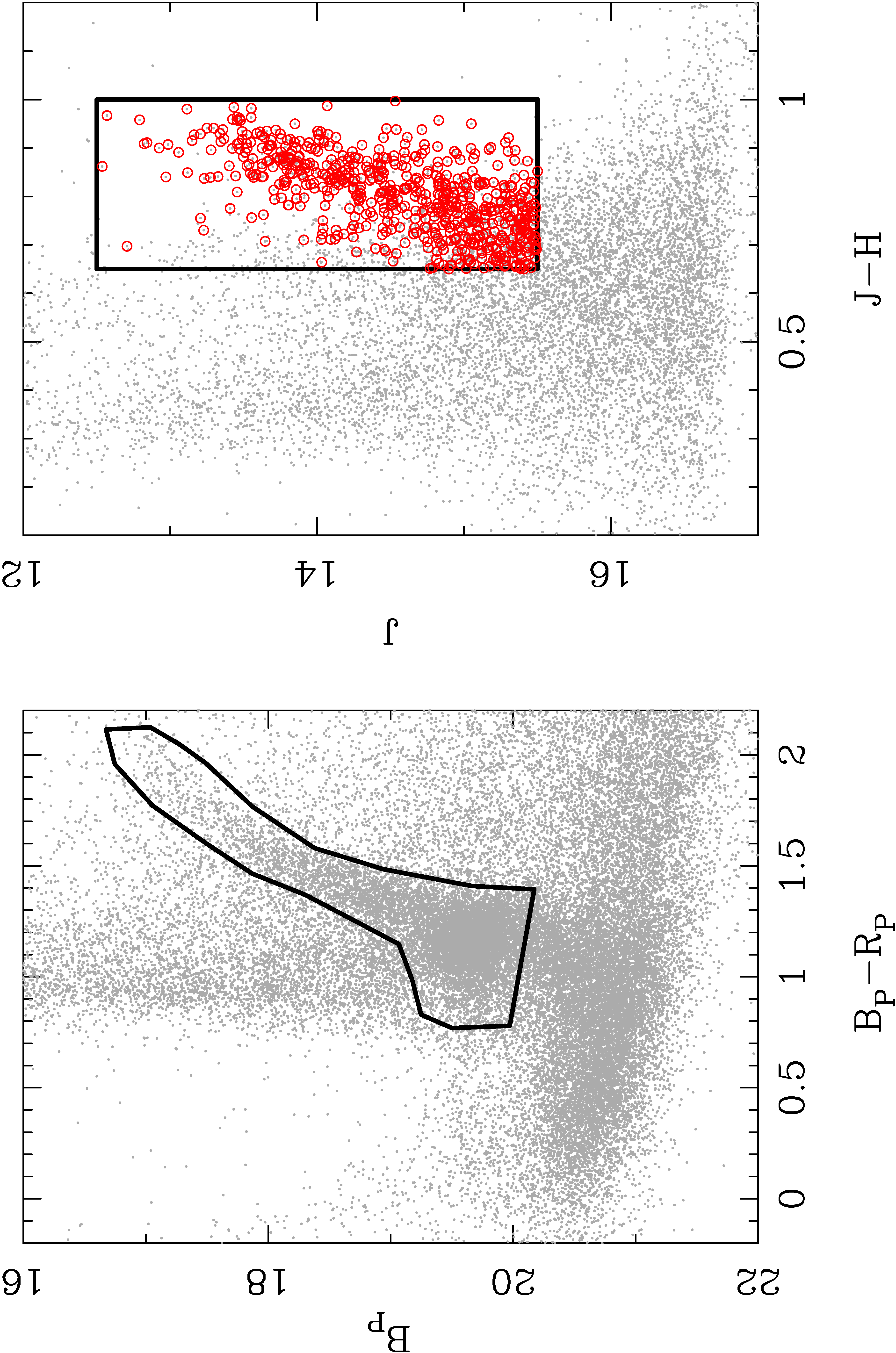}
\caption{CMD-selection of LMC field stars in Gaia, shown for the field of star 292. Left: using Gaia DR2 $B_P,R_P$ magnitudes; right: using 2MASS $J,H$ magnitudes. 
The black contours show our CMD selection. In the right plot, the red symbols highlight the stars within the CMD box {\bf and} with a parallax selection of $(\pi-2\sigma_{\pi}) \le 0.0$.}
\label{fig1}
\end{figure}

We extract from Gaia DR2 subfields of radius $1\arcdeg$ centered on each target star. We then identify LMC members within each subfield. Our first attempt was to select members from the Gaia DR2 color-magnitude diagram (CMD) in $B_P$ and $R_P$. We illustrate this selection in Figure \ref{fig1}, left panel, for the field of star 292. This field is the richest, and in addition to the well-represented giant branch and red clump, we also see a young population of stars at blue colors ($B_P-R_P \sim 0$). These are stars in the known young stellar association ICA76 located toward the Bridge; \citetalias{mb17} noted that star 292 belongs to this association. 
Unfortunately, the remaining five regions are in much less populated areas of the Cloud and as
such, foreground Milky Way stars overwhelm any selection made in the $B_{P}R_{P}$ CMD.  Compounding the problem, the Gaia DR2 proper-motion errors increase rapidly with magnitude. For these two reasons we have opted to use 2MASS \citet{skr06} to photometrically select LMC members. 
Gaia DR2 coordinates in each field were matched with 2MASS, using a tolerance of $0\farcs5$. The $J,H$ CMD in the field of 292 is shown in Fig.\ref{fig1}, right panel. Our CMD cut is a box with $12.5 \le J \le 15.5$ and $0.65 \le J-H \le 1.0$. The giant branch represented by M giants of intermediate age and metallicity is distinct from the foreground field in this $JH$ CMD (as evidenced early on by e.g., \citet{maj03}). These stars are further trimmed by distance, i.e., keeping those with parallaxes compatibile with zero at $2\sigma_{\pi}$ level.
Such stars are highlighted in Fig.\ref{fig1} (red symbols), with the parallax criterion effectively cleaning up the remaining foreground stars within the CMD box cut.

\begin{figure*}
\epsscale{1.15}
\plottwo{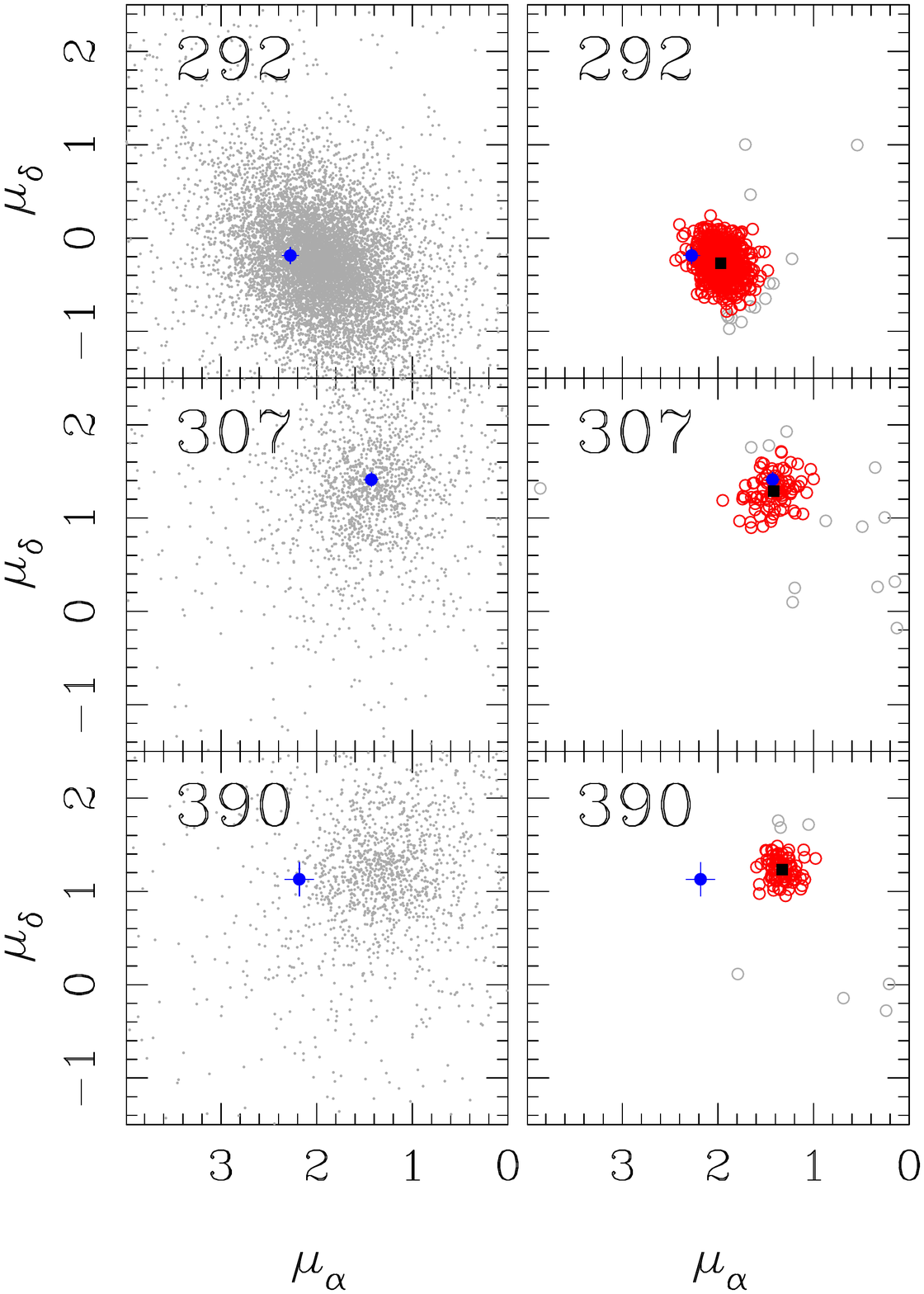}{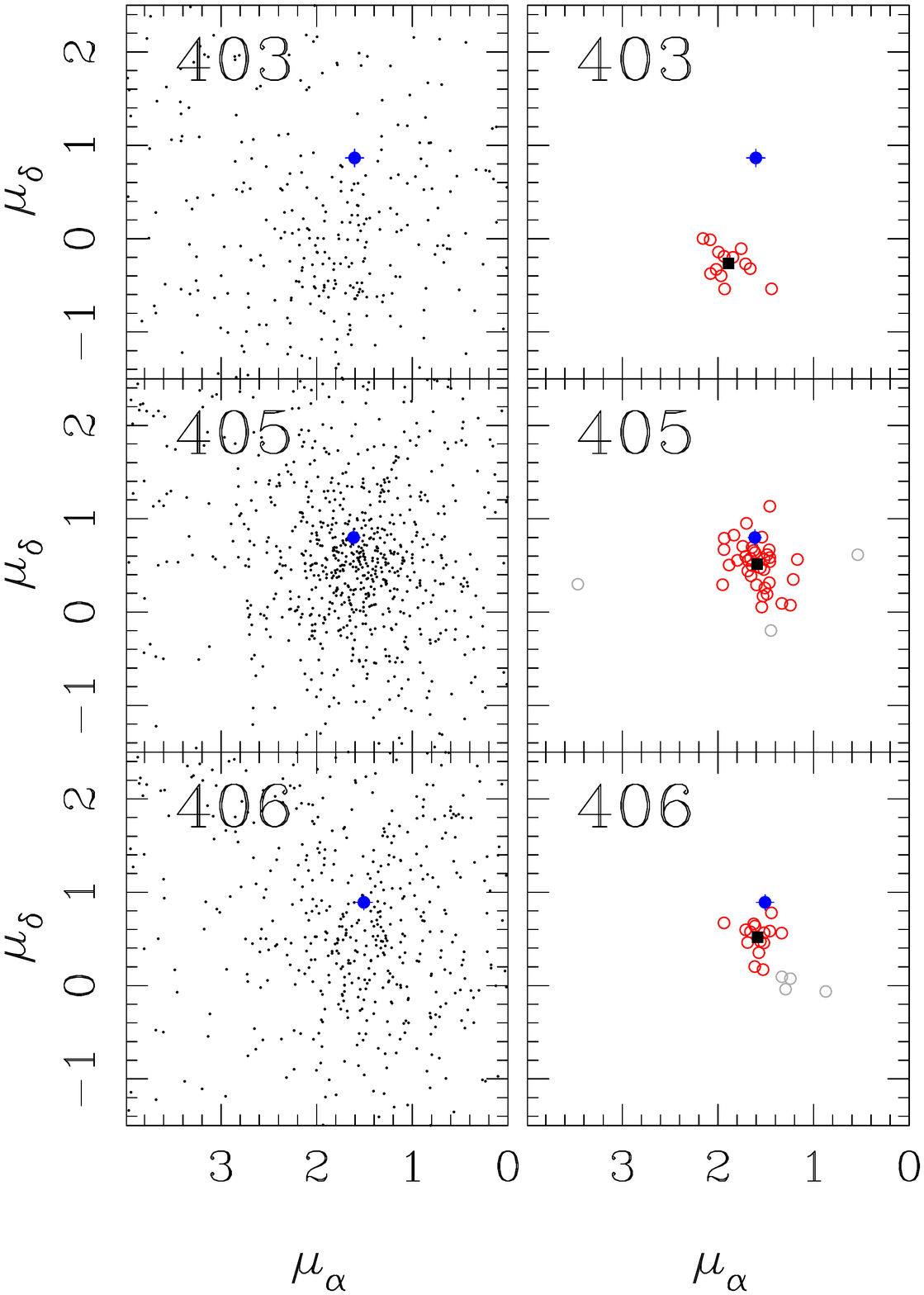}
\caption{Proper-motion diagram in the field of each star (as labeled). Left panels show the stars selected using the Gaia DR2 CMD plus parallax criterion (gray symbols), while the right panels show the stars selected using the 2MASS CMD plus parallax criterion (gray circles). The red symbols in the right panels highlight the stars selected to determine the mean proper motion within each field. The black filled square shows each mean; its uncertainty is smaller than the symbol size. The blue symbols show the proper motion of each of our young, blue stars.}
\label{fig2}
\end{figure*}

In Figure~\ref{fig2} we show the proper-motion diagrams within each subfield. 
Left panels show the $B_{P}R_{P}$-plus-parallax selected samples, while the right panels show the $JH$-plus-parallax selected samples. We purposely use the same plot limits for each subfield to illustrate the change in the mean motion from subfield to subfield, reflecting the rotation of the LMC disk. In all panels, the blue symbol with error bars shows the proper motion of the target, young star. The $JH$-selected samples exhibit much tighter proper-motion clumps compared to the $B_{P}R_{P}$-selected samples, confirming these are better-measured samples.

The mean proper motion within each subfield is determined from the $JH$-plus-parallax selected sample, after trimming proper-motion outliers by eye. Those stars used in each determination are highlighted in red in Fig.~\ref{fig2}. The resulting mean motion is represented with a black square. 
The values of these means, and the number of stars used in their determination, are listed in Tab.\ref{tab1}, columns 4, 5 and 3 respectively. Finally, the last two columns of Tab.\ref{tab1} show the proper-motion difference between the target star and the mean of the field. 
It is clear that stars 390 and 403 have proper motions significantly different from those of the local field, while stars 292, 307, 405, and 406 have proper motions within the dispersion of the local field. 
\begin{table*}[tbh]
\begin{center}
\caption{Proper motions and proper-motion differences between targets and field}
\label{tab1}
\begin{tabular}{c c r r r r r r r }
\hline
Gaia DR2 ID & ID & N$_{f}$ & 
\multicolumn{1}{c}{$<\mu_{\alpha}^f>$} & \multicolumn{1}{c}{$<\mu_{\delta}^f>$} & 
\multicolumn{1}{c}{$\mu_{\alpha}$} & \multicolumn{1}{c}{$\mu_{\delta}$} & 
\multicolumn{1}{c}{$\Delta \mu_{\alpha}$} & \multicolumn{1}{c}{$\Delta \mu_{\delta}$}  \\
\hline
4629325831365569408 & 292 & 620 & $1.971(0.006)$ & $-0.272(0.007)$ & $2.275(0.086)$ & $-0.187(0.088)$ & $0.304(0.086) $& $0.085(0.088)$ \\ 
5263888695091870976 & 307 & 82 & $1.418(0.020)$ & $1.288(0.020)$ & $1.429(0.070)$ & $1.412(0.077)$ & $0.011(0.073) $& $0.124(0.080)$ \\ 
5281835782874813952 & 390 & 81 & $1.329(0.013)$ & $1.231(0.014)$ & $2.182(0.151)$ & $1.129(0.180)$ & $0.853(0.152) $& $-0.102(0.181)$ \\ 
4774221707057337088 & 403 & 13 & $1.890(0.056)$ & $-0.236(0.049)$ & $1.603(0.096)$ & $0.864(0.095)$ & $-0.287(0.111) $& $1.127(0.107)$ \\ 
4764998110170179328 & 405 & 36 & $1.589(0.052)$ & $0.515(0.041)$ & $1.614(0.068)$ & $0.798(0.085)$ & $0.025(0.086) $& $0.283(0.094)$ \\ 
5495816929075206912 & 406 & 15 & $1.584(0.036)$ & $0.516(0.044)$ & $1.507(0.091)$ & $0.890(0.081)$ & $-0.077(0.098) $& $0.374(0.092)$ \\ 
\hline
\end{tabular}
\end{center}
\end{table*}

Proper-motion differences are multiplied by a fixed 50-kpc distance, in agreement with the LMC's distance modulus of 18.49 \citep{pie13}, 
yielding velocities in the plane of the sky. 
The total tangential-velocity differences, and their uncertainties\footnote{As calculated, the uncertainty in $\Delta V_{T}$ is dominated by the Gaia DR2 proper-motion uncertainty of each target star.
While the known correlations between the $\alpha$ and $\delta$ components of the Gaia $\mu$ measures \citep{lind18} will affect the final uncertainty in 
$\Delta V_{T}$, in actuality the effect of the correlation amounts to $\sim 1$ km s$^{-1}$ or less for these stars.
This is negligible relative to the overall uncertainty values of $\sim 20-30$ km s$^{-1}$ (see Table \ref{tab2}) and has been ignored.}
are listed in Table \ref{tab2} as $\Delta V_{T}$.
Ages, as derived by \citetalias{mb17}, are also listed in Tab. \ref{tab2} along with their estimated uncertainties. 
These are combined to calculate lifetime tangential-travel distances that are also given in Tab. \ref{tab2}, in both degrees and kpc. 
For comparison, we also list in the last column of Tab. \ref{tab2}, the line-of-sight velocity difference $\Delta V_{los}$ between the target star and the 
prediction of an LMC disk model as determined in \citetalias{mb17}. $\Delta V_{los}$ is not used in the determination of the travel distance.

There are several implicit assumptions to our analysis that deserve discussion.  First,
in order to convert proper-motion differences to velocity differences, it is assumed the
young target stars and the (angularly) nearby LMC stars are at the same mean distance.
This is reasonable given that the young star and the M giants are likely to both belong to the
disk of the LMC. (Regardless, we are attempting to ascertain if the two are comoving and it
is highly unlikely that a combination of discordant distances and tangential velocities would
conspire to yield such a small proper-motion difference as is seen in four of the six cases.)
Second, we assume that the mean motion of M giants within each field is representative of the
local LMC disk motion. Third, we assume that while velocity gradients across each $2\arcdeg$-diameter area may distort the proper-motion distribution, the location of the target star at the spatial center of the field ensures that such gradients should not affect the mean motion.

Only stars 390 and 403 show velocity differences in excess of 200 km~s$^{-1}$, with
corresponding large travel distances of $\sim 6-7$ kpc. The remaining stars have velocity differences $< 100$ km~s$^{-1}$, and travel distances of the order of 1 kpc. A marginal exception is star 406, with a travel distance of 4 kpc, due to its having the largest age in our sample. However, its near proximity to star 405 and their similar proper-motion differences (i.e., velocity vectors) hint that they were possibly formed in the same local area. 
We note star 292 is a known member of a nearby young association. As such, this star serves as an example of an in-situ formation referred to the local velocity field of an intermediate-age M-giant population.
\begin{table}
\begin{center}
\caption{Velocity and travel distance}
\label{tab2}
\begin{tabular}{c r r r r r}
\hline
ID  & \multicolumn{1}{c}{$\Delta V_{T}$} & \multicolumn{1}{c}{Age} & \multicolumn{2}{c}{Dist} & \multicolumn{1}{c}{$\Delta V_{los}$}\\
   & \multicolumn{1}{c}{(km~s$^{-1}$)} & \multicolumn{1}{c}{(Myr)} & \multicolumn{1}{c}{($\arcdeg$)} & \multicolumn{1}{c}{(kpc)}  & 
\multicolumn{1}{c}{(km~s$^{-1}$)} \\
\hline
292  & $75(21)$  & $11(06)$   & 1.0(0.6) & $0.8(0.5)$ & $-52(9)$\\
307  & $29(19)$  & $18(05)$   & 0.6(0.2) & $0.5(0.4)$ & $14(8)$\\
390  & $204(36)$ & $35(08)$   & 8.4(2.4) & $7.1(2.1)$ & $-28(7)$\\
403  & $276(25)$ & $20(05)$   & 6.5(1.7) & $5.5(1.5)$ & $-48(7)$\\
405  & $67(22)$  & $20(05)$   & 1.6(0.5) & $1.3(0.6)$ & $-21(5)$\\
406  & $90(22)$  & $45(10)$   & 4.8(1.3) & $4.1(1.3)$ & $40(8)$\\
\hline
\end{tabular}
\end{center}
\end{table}

\begin{figure}
\includegraphics[width=0.32\textwidth,angle=-90]{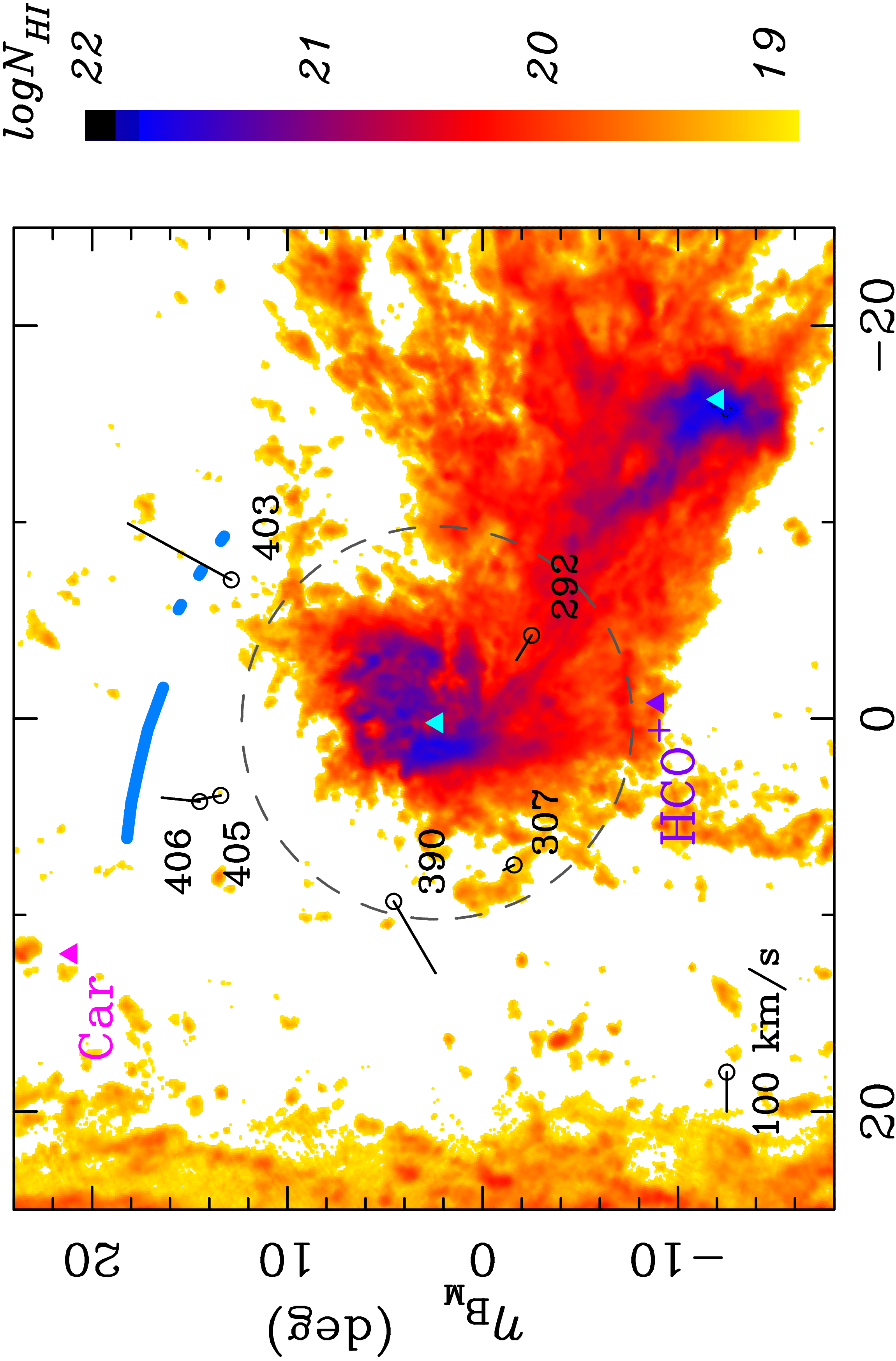}
\par
\vspace{0.5cm}
\includegraphics[width=0.375\textwidth,angle=-90]{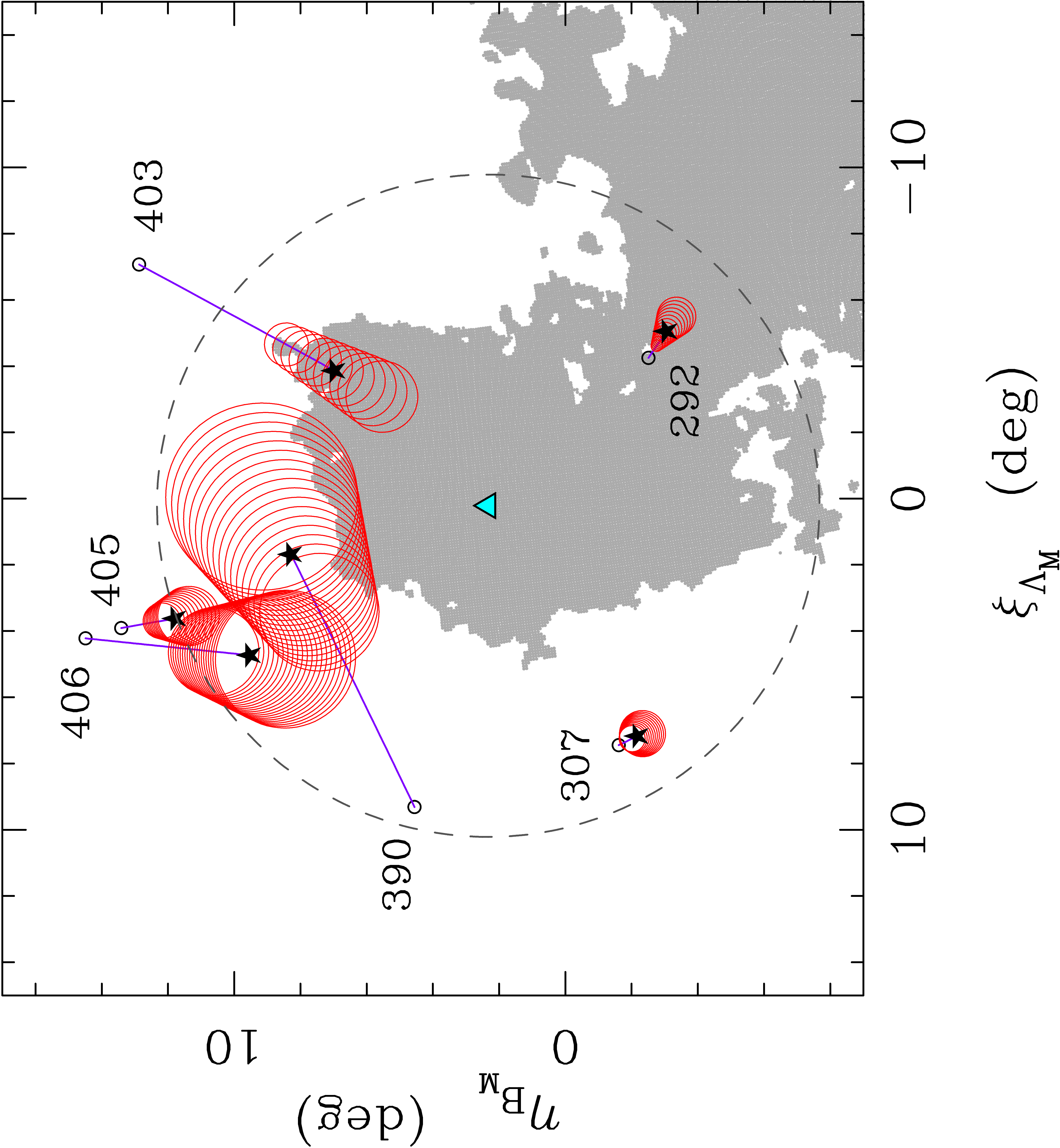}
\caption{Spatial distribution of our young stars, as a gnomonic projection of Magellanic coordinates. {\bf Top:}
The color map shows the H~I column density from the GASS survey with velocities $V_{lsr} = 100$ to $450$ km~s$^{-1}$. 
The centers of the LMC and SMC are marked with blue triangles.
Other objects of interest are marked and labeled: the center of the Carina dwarf spheroidal galaxy, the location of HCO$^{+}$ detection \citep{murray15}, 
and the arc/tidal feature (pale blue) identified in the Dark Energy Survey by \citet{mackey16}. 
Velocity differences in the plane of the sky are shown for each of our target stars. 
This panel is a modified version of Fig. 7 in \citetalias{mb17}.
{\bf Bottom:} Current-day positions of our target stars (open circles) and their likely positions of formation (star symbols) as
derived by projecting their motions backward in time. For each star the series of error ellipses (in red)
indicates 1-Myr steps in time, extending $\pm 1 \sigma$ the estimated age of the star.  The gray background
indicates regions where H~I density exceeds $10^{20}$cm$^{-2}$. In both panels the dashed circle indicates a $10\arcdeg$ radius from the LMC's center.}
\label{fig4}
\end{figure}

In Figure \ref{fig4} we show the distribution of our target stars in the plane of the sky. We use Magellanic coordinates \citep{nide08} in a gnomonic projection centered on the LMC. The top panel shows the H~I column density map from the GASS survey \citep{mcc09,kal15} with $V_{lsr} = 100 - 450$ km ~s$^{-1}$. For each star, the velocity-difference vectors are indicated. The two stars with velocity differences in excess of 200 km~s$^{-1}$ are moving away from
the inner regions of the LMC disk. Thus, their likely origin is in the denser parts of the disk, having been ejected toward the outskirts by dynamical interactions of the type described by \citet{boub17}. Their velocities in excess of 200 km~s$^{-1}$ at current radii indicate that they escape the LMC \citep[see e.g., Fig. 4 in][]{boub17}.
The remaining four stars appear to have formed locally within a few degrees of their present location. 


We illustrate the likely origin of the six stars in the bottom panel of Figure~\ref{fig4}.  The
difference in proper motion between each star and the mean motion of its nearby LMC members is
combined with the star's age estimate to predict the location of the star when it formed by
simply propagating backward in time by the
star's age.  In the figure, error ellipses are drawn at 1 Myr intervals around the star's 
estimated age, $\pm 1\sigma$ the age uncertainty.  Thus, for each star, the complex of error
ellipses represents its likely place of origin.  For reference, the region with H~I column
density exceeding $10^{20}$cm$^{-2}$ is shown in gray.  Evidently, stars 390 and 403 formed
in regions with higher gas density, compared to their present location, while 
stars 307, 405, and 406 likely formed in low-density regions. 
Of course, this simple approach does not explicitly include the LMC's potential when tracing back in 
time each star's position.  
Nonetheless, for the slow moving stars this differential approach is appropriate and allows us to reach a definite conclusion regarding their origin.
For the two fast moving stars the approach is admittedly simplified, but should still indicate the rough direction and amount of offset of each star's origin 
location relative to its nearby LMC neighbors.  
A more rigorous orbit integration might possibly improve the estimate, but considering the size of the uncertainties in age and transverse velocity,
we do not think it is warranted at this time.

It is necessary to explore the possibility that stars 390 and 403 were expelled from the Milky Way. 
To do so, we integrate back in time the orbits of these two stars in an analytic
3-component Galactic potential \citep{jsh95}, using the distance moduli derived in \citetalias{mb17}, and ignoring the LMC. We find that the pericenter
of star 390 is 49 kpc, reached some 17 Myr ago, i.e., within its current age range, compared to its current Galactocentric distance of 50 kpc. 
For star 403, the predicted pericenter is 52 kpc some 39 Myr ago, i.e., beyond its age range, while its current Galactocentric distance is 54 kpc. 
Young stars with pericenters of $\sim 50$ kpc are unlikely to have originated in our Galaxy.  
They more likely escaped from a more gas-rich region of the LMC, and are analagous to high-velocity
star HVS3 that was recently confirmed to have a Magellanic origin \citep{er18}. 

\section{Summary} \label{sec:sum}
We use Gaia DR2 data to confirm the origin of six young stars located in the outskirts of the LMC. We find that four stars have low velocities. Combining this with age estimates derived in an earlier study \citepalias{mb17}, the four stars must have been born within $1\arcdeg$ to $\sim 5\arcdeg$ of their current location. Three of these stars do not belong to any known young association and have formed in very low H~I density regions, in the periphery of the Cloud. It is conceivable that the recent ($\sim 200$ Myr) collision between the Small Magellanic Cloud and the LMC could have triggered star formation in the far outskirts of the LMC's disk \citepalias[see][and references therein]{mb17}.
The remaining two stars have velocity differences in excess of 200 km~s$^{-1}$ and in directions roughly outward from the LMC, indicating their origin is consistent with being runaways from the inner LMC.
\\ \\
DIC and TMG were supported, in part, by NASA grant 80NSSC18K0422.
DIC is grateful to the Vatican Observatory Summer School where she was supported as a lecturer during revision of this manuscript, (and where her lab students were given the opportunity to replicate the results of this study!). 
LZ acknowledges support by the National Science Foundation of China  grants 11773033 and 11390371/2.
VIK acknowledges support by grant 3.858.2017/4.6 of the Ministry of Education and Science (Russia).
This work has made use of data from the European Space Agency (ESA) mission
{\it Gaia} (\url{https://www.cosmos.esa.int/gaia}), processed by the {\it Gaia}
Data Processing and Analysis Consortium (DPAC,
\url{https://www.cosmos.esa.int/web/gaia/dpac/consortium}). Funding for the DPAC
has been provided by national institutions, in particular the institutions
participating in the {\it Gaia} Multilateral Agreement.
This publication makes use of data products from the Two Micron All Sky Survey, which is a joint project of the University of Massachusetts and the Infrared Processing and Analysis Center/California Institute of Technology, funded by the NASA and the NSF.

\end{document}